# Practical GHz single-cavity all-fiber dual-comb laser for high-speed spectroscopy

Lin Ling[1†], Wei Lin[1†], Zhaoheng Liang[1†], Minjie Pan[1], Chiyi Wei[1], Xuewen Chen[1], Yang Yang[1], Zhijin Xiong[1], Yuankai Guo[1], Xiaoming Wei[1,2*], and Zhongmin Yang[1,2,3*]

[1]School of Physics and Optoelectronics, South China University of Technology, 381 Wushan Road, Guangzhou 510640, China.

[2]Guangdong Engineering Technology Research and Development Center of Special Optical Fiber Materials and Devices; Guangdong Provincial Key Laboratory of Fiber Laser Materials and Applied Techniques, School of Materials Science and Engineering, South China University of Technology, Guangzhou 510641, China.

[3]Research Institute of Future Technology, South China Normal University, Guangzhou 510006, China.

[†]These authors equally contribute to this work.

[*]Correspondence should be addressed to X.M.W. (xmwei@scut.edu.cn) or Z.M.Y. (yangzm@scut.edu.cn).

## Abstract

Dual-comb spectroscopy (DCS) with few-GHz tooth spacing that provides the optimal trade-off between spectral resolution and refresh rate is a powerful tool for measuring and analyzing rapidly evolving transient events. Despite such an exciting opportunity, existing technologies compromise either the spectral resolution or refresh rate, leaving few-GHz DCS with robust design largely unmet for frontier applications. In this work, we demonstrate a novel GHz DCS by exploring the multimode interference-mediated spectral filtering effect in an all-fiber ultrashort cavity configuration. The GHz single-cavity all-fiber dual-comb source is seeded by a dual-wavelength mode-locked fiber laser operating at fundamental repetition rates of about 1.0 GHz differing by 148 kHz, which has an excellent stability in the free-running state that the Allan deviation is only 101.7 mHz for an average time of 1 second. Thanks to the large repetition rate difference between the asynchronous dichromatic pulse trains, the GHz DCS enables a refresh time as short as 6.75 μs, making it promising for studying nonrepeatable transient phenomena in real time. To this end, the practicality of the present GHz DCS is validated by successfully capturing the 'shock waves' of balloon and firecracker explosions outdoors. This GHz single-cavity all-fiber dual-comb system promises a noteworthy improvement in acquisition speed and reliability without sacrificing measurement accuracy, anticipated as a practical tool for high-speed applications.

## Introduction

Optical frequency comb (OFC) has been recognized as a revolutionary tool with incomparable accuracy for precision measurement [1-3]. In particular, in the fashion of exploiting two OFCs with slightly different repetition rates, dual-comb spectroscopy (DCS) cannot only rapidly interrogate the spectrum without mechanical inertia that is typically required for conventional technologies, but also further increase the measurement accuracy by leveraging heterodyne detection [4-9], making it powerful in high-speed spectroscopy [10], accuracy metrology [11], hyperspectral optical microscopy [12], high-sensitivity optical sensing [13], etc. To enable faster acquisition speed, an increasing repetition rate difference is essential, for which the asynchronous pulse trains must operate at high repetition rates. To this end, great efforts have been dedicated to the generation of OFCs at 10's to 100's GHz, particularly chip-based microresonator combs [14], as well as quantum cascade laser combs [15] and electro-optic modulation combs [16]. Although prior works are capable of DCS with temporal resolutions of microseconds to nanoseconds [17-19], it is noticed that DCS at few-GHz can provide the optimal trade-off between spectral resolution and refresh rate for measuring sparse events, which however is challenging for chip-based OFC technologies [20,21].

As an alternative resolution, the mode-locked fiber laser (MLFL) has been intensively explored for generating OFC as its compactness and reliability, however with fundamental repetition rates long being limited to 10's MHz, leaving the few-GHz

MLFL-based DCS unmet. Recently, several works have reported MLFL-based OFCs with GHz-level fundamental repetition rates [22-25]. However, traditional MLFL-based GHz DCS requires additional sophisticated and cumbersome configurations to achieve high coherence and good stability between two independent MLFLs [26,27], which doubtlessly increases the complexity and thus limits practical applications. To circumvent this issue, single-cavity dual-comb MLFL is a promising solution [28-33], as therein the asynchronous pulse trains share the same laser cavity that promises high reliability and coherence. Despite this exciting potential, a practical GHz single-cavity all-fiber dual-comb laser has yet to be demonstrated by exploring high-gain active fiber for shortening the laser cavity and new mechanisms for generating asynchronous pulse trains at GHz.

In this paper, we present a practical single-cavity dual-comb source by exploring dual-wavelength MLFL with GHz-level fundamental repetition rates based on a new mechanism of multimode interference (MMI)-mediated spectral filtering effect in the ultrashort fiber cavity. The new single-cavity all-fiber dual-comb laser enables the optimal performance merit for high-speed DCS applications, including fundamental repetition rates of about 1.0 GHz and a repetition rate difference of 148 kHz. In the free-running state, the fluctuation of the repetition rate difference is measured to be only 101.7 mHz (Allan deviation) for an average time of 1 second. The GHz single-cavity all-fiber dual-comb system is integrated to validate its practical applications outdoors, and transient processes of balloon and firecracker explosions are successfully captured and analyzed.

## Results

**Figure 1** conceptually illustrates the principle of the single-cavity all-fiber dual-comb source that is generated from a dual-wavelength MLFL with GHz-level fundamental repetition rates. The MLFL leverages the passive mode-locking technology in a standard Fabry-Pérot (FP) cavity, which is implemented by using a semiconductor saturable absorber mirror (SESAM) as one reflector, while a dielectric film (DF) as the other one. The formation of asynchronous dichromatic pulses is associated with an all-fiber spectral filter in a sandwich structure consisting of passive single-mode fiber (SMF) and active few-mode gain fiber (FMGF), i.e., an equivalent SMF-FMGF-SMF structure in a single round-trip pass (middle left insets of **Fig. 1**), which results in the MMI between the fundamental mode (i.e., $LP_{01}$) and high-order modes (HOMs, i.e., $LP_{11}$ in this case). More specifically, the HOM can be excited when the fundamental-mode light field in the SMF enters the FMGF, primarily the $LP_{11}$ mode regarding a normalized frequency of 2.809. The pulse in the HOM propagates at a different speed than that of the fundamental mode due to the intermodal group velocity difference. The MMI-mediated spectral filtering effect occurs when the multimode light fields are coupled back into the SMF, and thereby gives rise to asynchronous dichromatic pulses because of the chromatic dispersion. It is worth noting that sufficient gain provided by the short FMGF is required for generating passively mode-locked pulse trains with GHz-level fundamental repetition rates, such that a heavily doped FMGF with high gain is utilized as the gain medium. More details are provided in **Methods**, as well as **Supplementary Notes 1** and **2**.

For dual-comb generation, the spectral overlap between the asynchronous dichromatic pulses is further imparted by coherently spectral broadening (bottom panels of **Fig. 1**). Then, the two spectrum-overlapped optical combs with a frequency offset (i.e., repetition rate difference) are down converted into radio frequency (RF) comb through heterodyne detection, and the mapping factor is determined by the ratio of fundamental repetition rate to frequency offset. The high-speed spectroscopic information is extracted by applying the Fourier transform to the temporal interferograms (right column of **Fig. 1**). Thanks to high fundamental repetition rates of $f > 1$ GHz that enable a substantial frequency offset of $\Delta f > 100$ kHz, the refresh time of the DCS is less than 10 μs.

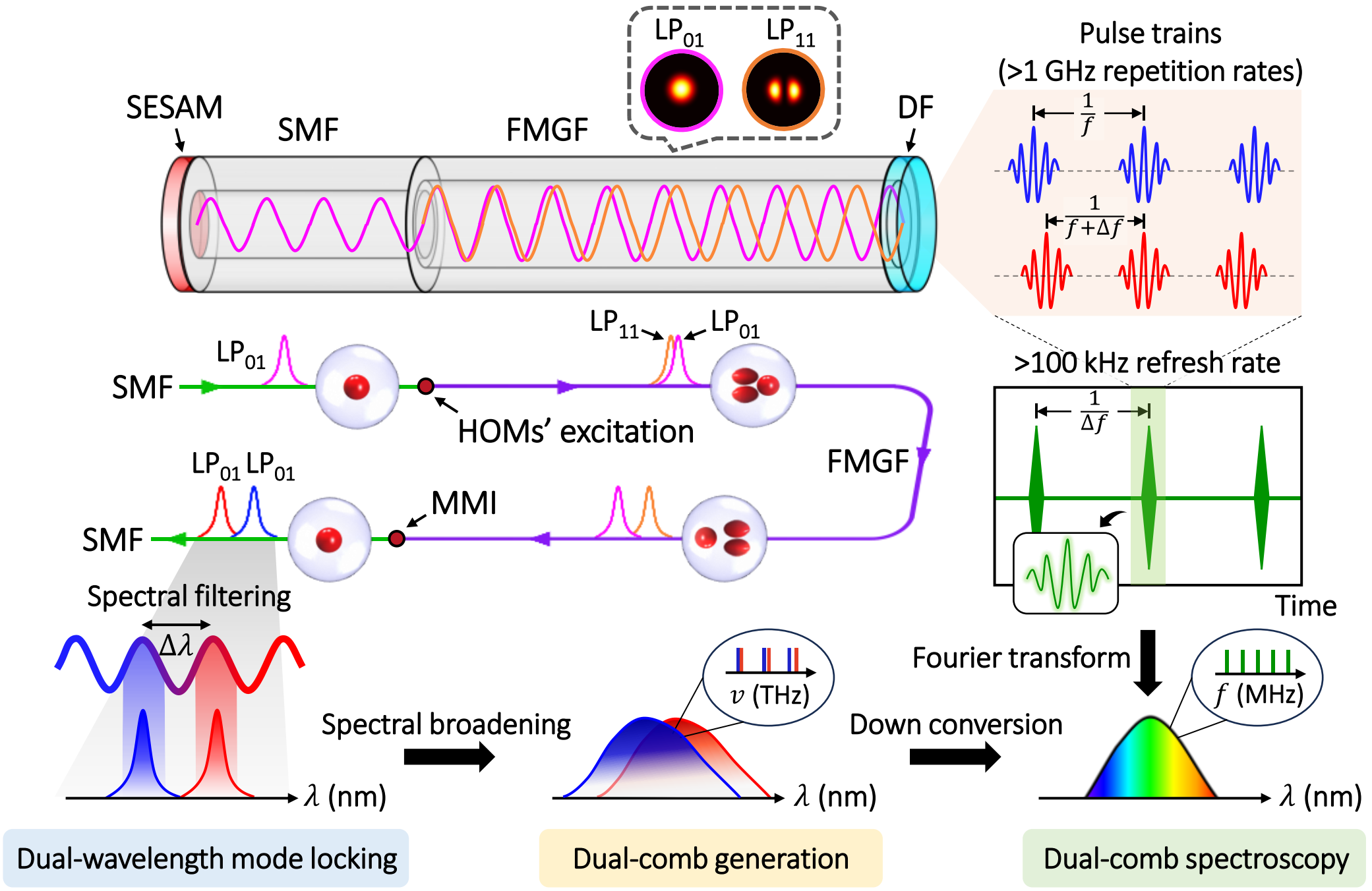

**Fig. 1 | Principle of GHz single-cavity all-fiber dual-comb laser.** The dual-comb generation is seeded by a free-running dual-wavelength mode-locked fiber laser (MLFL) with GHz-level fundamental repetition rates ($f$). In the laser cavity, a single-mode fiber (SMF) is spliced with a few-mode gain fiber (FMGF), and they are sandwiched by a semiconductor saturable absorber mirror (SESAM) and a dielectric film (DF) that serve as the cavity reflectors. The coupling between SMF and FMGF gives rise to a multimode interference (MMI)-mediated spectral filtering effect between the fundamental and high-order modes (HOMs), like $LP_{01}$ and $LP_{11}$ in this case (middle left insets), and facilitates the generation of dichromatic mode-locked pulse trains (top right insets). According to the group velocity difference between the two wavelengths, the pulse trains are asynchronous with a repetition rate difference ($\Delta f$) on the order of 100 kHz. The optical spectra of asynchronous dichromatic pulses are then broadened for spectral overlap, i.e., generating the GHz dual-comb source (bottom panels). For dual-comb spectroscopy (DCS), the GHz dual-comb signal is down converted into a radio frequency (RF) comb. The temporal interferograms generated from the beating of the two asynchronous dichromatic pulse trains are further processed by Fourier transform to extract the high-speed spectroscopic information (right panels).

## Mechanism and implementation of MMI-mediated dual-wavelength mode locking

To understand how the MMI-mediated spectral filtering effect assists the generation of asynchronous dichromatic pulse trains with GHz-level fundamental repetition rates in the ultrashort fiber cavity, we first investigate the characteristics of the MMI-mediated spectral filter. In the FP laser cavity, the coupling between the passive SMF and Yb-doped FMGF yields the SMF-FMGF-SMF spectral filter in a single round-trip pass (inset of **Fig. 2a**). By performing the modal analysis of the FMGF (**Supplementary Note 2**), the wavelength spacing $\Delta\lambda$ resulted from the MMI-mediated spectral filtering effect is governed by

$$\Delta\lambda = \left|\frac{\lambda^2}{2Lc\delta\beta_1}\right|, \tag{1}$$

where $\lambda$ is the operating wavelength. $\delta\beta_1$ represents the group velocity difference between transverse modes $LP_{01}$ and $LP_{11}$. $L$ is the length of the FMGF, and $c$ is the speed of light. The calculated result of **Eq. (1)** is illustrated in **Fig. 2a**, while the experimentally measured transmission curve is shown in **Fig. 2b**. As can be observed, the comb-like transmission spectrum exhibits a wavelength spacing of 6.28 nm, which is in good consistency with the theoretical calculation for an operating wavelength of 1059 nm (i.e., 6.31 nm). More details are provided in **Supplementary Note 3**.

Generating asynchronous dichromatic pulses in ultrashort fiber cavities with limited gain by exploring the MMI-mediated spectral filtering effect is challenging, although it has been successfully demonstrated in long-fiber cavities with far lower fundamental repetition rates [34,35]. The noteworthy challenges include: (1) the accompanying gain filtering effect may affect the dual-wavelength mode locking at GHz-level fundamental repetition rates [36]; (2) the cavity-induced soliton trapping effect also intrinsically hinders the asynchrony of dichromatic pulses [37]. As a result, the MMI-mediated spectral filter needs to be carefully designed for successfully generating asynchronous dichromatic pulses. Here we first perform theoretical studies by investigating a revised model, wherein the gain filtering is modeled by a set of propagation-rate equations (**Methods**). With varying wavelength spacing of the MMI-mediated spectral filter, here increasing from 3.6 to 8.3 nm, the operation of the MLFL evolves from the bound state with synchronous pulses to the dual-comb state with asynchronous pulses (**Fig. S4c**). The underlying physics can be understood as: with a relatively small difference in the operating wavelengths (e.g., $\Delta\lambda$ = 4.2 nm, left panels of **Fig. 2c**), the dichromatic pulses can be bound with each other as a synchronous unit owing to the intracavity trapping mechanism dominated by the SESAM (**Supplementary Note 4.1**). In this case, the group velocity difference provided by the small net dispersion is limited in such an ultrashort fiber cavity. As the wavelength spacing exceeds 4.7 nm, the cavity-induced trapping effect is insufficient for compensating the chromatic dispersion-induced walk-off, thus resulting in the generation of asynchronous dichromatic pulses (e.g., $\Delta\lambda$ = 6.3 nm, right panels of **Fig. 2c**). More details about the transition from bound state to dual-comb state are discussed in **Supplementary Note 4.1**.

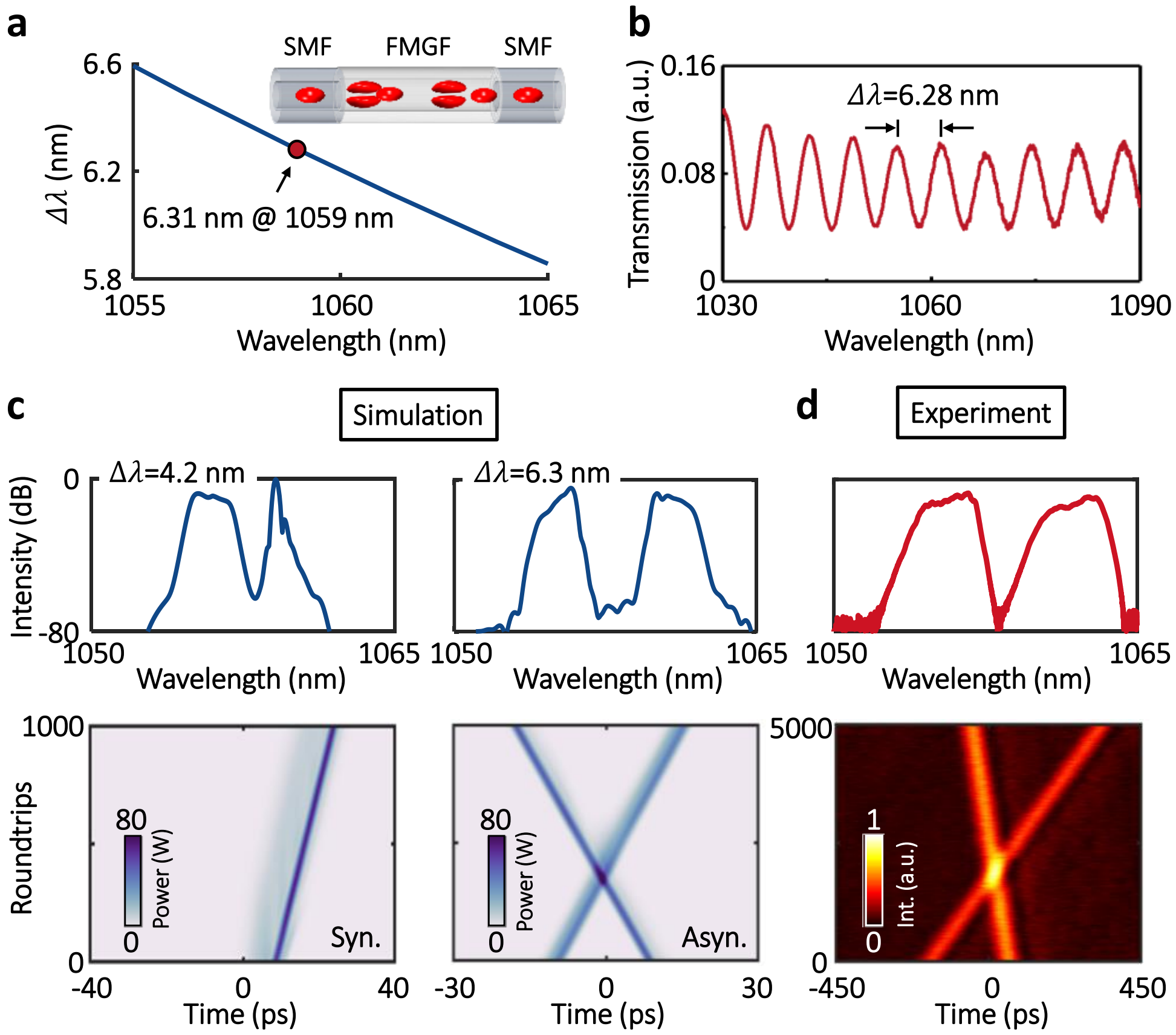

**Fig. 2 | Implementation of MMI-mediated dual-wavelength mode locking at GHz. a.** Calculated wavelength spacing $\Delta\lambda$ of the MMI-mediated spectral filtering effect. Inset shows the laser cavity that has an equivalent SMF-FMGF-SMF configuration. **b.** Measured intracavity transmission curve. **c.** Simulated optical spectra (top) and temporal evolutions (bottom) for synchronous (left) and asynchronous (right) mode-locked pulses. **d.** Experimental optical spectrum (top) and temporal evolution (bottom) of the dual-wavelength MLFL.

In experimental implementation, we successfully achieve asynchronous dichromatic pulse trains with fundamental repetition rates of about 1.0 GHz. As shown in **Fig. 2d**, wherein the optical spectrum and temporal evolution closely match with the simulation results. The center wavelengths of the asynchronous dichromatic pulses are located at around 1056 nm and 1062 nm, with 3-dB bandwidths of 2.12 nm and 1.54 nm, respectively. It is also noted that the inevitable intracavity collision between the asynchronous dichromatic pulses may potentially deteriorate the performance of the DCS. To this end, the intracavity collision is studied from both numerical and experimental perspectives (**Supplementary Note 4.2**). It shows that the energy exchange is limited throughout the collision process, and only minor change is observed on the edges of the optical spectra. More details about dual-wavelength MLFL are provided in **Supplementary Note 5**.

**Frequency stability of asynchronous dichromatic pulse trains**

We further experimentally examine the stability of asynchronous dichromatic pulse trains to identify the practicality for DCS applications. **Figure 3a** shows the RF spectrum with a resolution bandwidth (RBW) of 10 Hz in a frequency span of 2 MHz, which indicates the fundamental repetition rates of about 1.093 GHz, matching well with the effective length of the laser cavity (i.e., 9.3 cm). As operating in a normal dispersion regime, the asynchrony of the dichromatic pulse trains is also verified by the difference in their fundamental repetition rates, i.e., 1.093118 GHz ($f_1$) and 1.093266 GHz ($f_2$) for center wavelengths of 1056 nm and 1062 nm, respectively. The main RF signal exhibits a signal-to-noise ratio of > 70 dB, confirming the excellent short-term stability of the dual-wavelength mode locking. It is also noted that a series of symmetrical subsidiary peaks presented in the RF spectrum are resulted from the beating between the asynchronous dichromatic pulse trains, as shown in the inset of **Fig. 3b**, wherein it illustrates the temporal evolution of the pulse trains. The overlap of the asynchronous dichromatic pulse trains manifests distinguishing characteristics in different time scales: (1) each pulse train has a relatively uniform intensity in the short time scale (**Fig. 3b**), and it clearly shows the temporal separation between the pulses of about 914 ps, which agrees well with the RF measurement; (2) there exists periodic beat notes at every 6.75 μs in the long time scale (inset of **Fig. 3b**), consistent with their repetition rate difference $\Delta f = f_2 - f_1$ of 148 kHz.

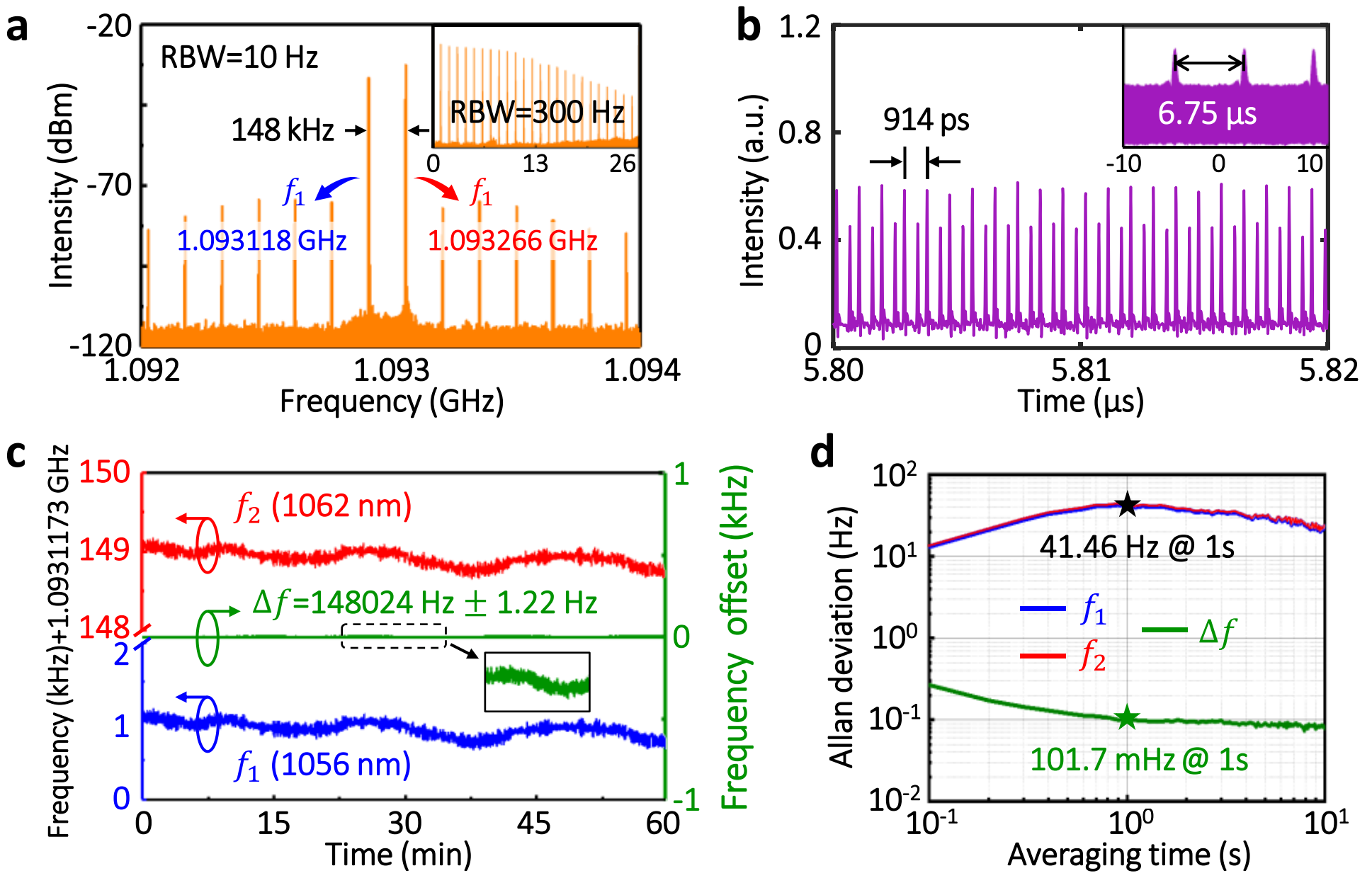

**Fig. 3 | Frequency stability of asynchronous dichromatic pulse trains at GHz. a.** RF spectrum of the asynchronous dichromatic pulse trains in a frequency span of 2 MHz with a resolution bandwidth (RBW) of 10 Hz. Inset shows the RF spectrum in a wider span with an RBW of 300 Hz. **b.** Oscilloscopic trace of the asynchronous dichromatic pulse trains in a time window of 20 ns. Inset shows the pulse trains in a wider time window. **c.** Stability measurements of the asynchronous pulse trains at repetition rates $f_1$ (blue), $f_2$ (red) and their difference $\Delta f$ (green) for 60 minutes. **d.** Allan deviations calculated from **c**.

As shown in **Fig. 3c**, the repetition rates $f_1$ (blue) and $f_2$ (red) of the asynchronous dichromatic pulse trains, as well as their difference $\Delta f$ (green), are recorded every 100 ms for 60 minutes in the free-running state. Both $f_1$ and $f_2$ exhibit obvious fluctuation with a standard deviation of about 90 Hz, and they follow similar increasing or decreasing trend of frequency change, such that the fluctuation of $\Delta f$ suppressed by two orders of magnitude, i.e., only 344 mHz. **Figure 3d** presents the Allan deviations of $f_1$, $f_2$ and $\Delta f$, wherein $f_1$ and $f_2$ have indistinguishable Allan deviations, i.e., about 41.46 Hz for an averaging time of 1 second, while it is only 101.7 mHz for $\Delta f$. The excellent stability performance of $\Delta f$ can be attributed to the physical superiority that the asynchronous dichromatic pulses are propagating in the same laser cavity and share identical influences of environmental disturbances. This dual-wavelength MLFL with such good stability of $\Delta f$ in the free-running state can promise excellent DCS performance with simple and reliable design, such that the complexity of dual-comb system compared to the conventional methods can be significantly reduced [38,39]. Its long-term stability as well as the noise performance is also characterized in **Supplementary Note 6**.

## Dual-comb generation

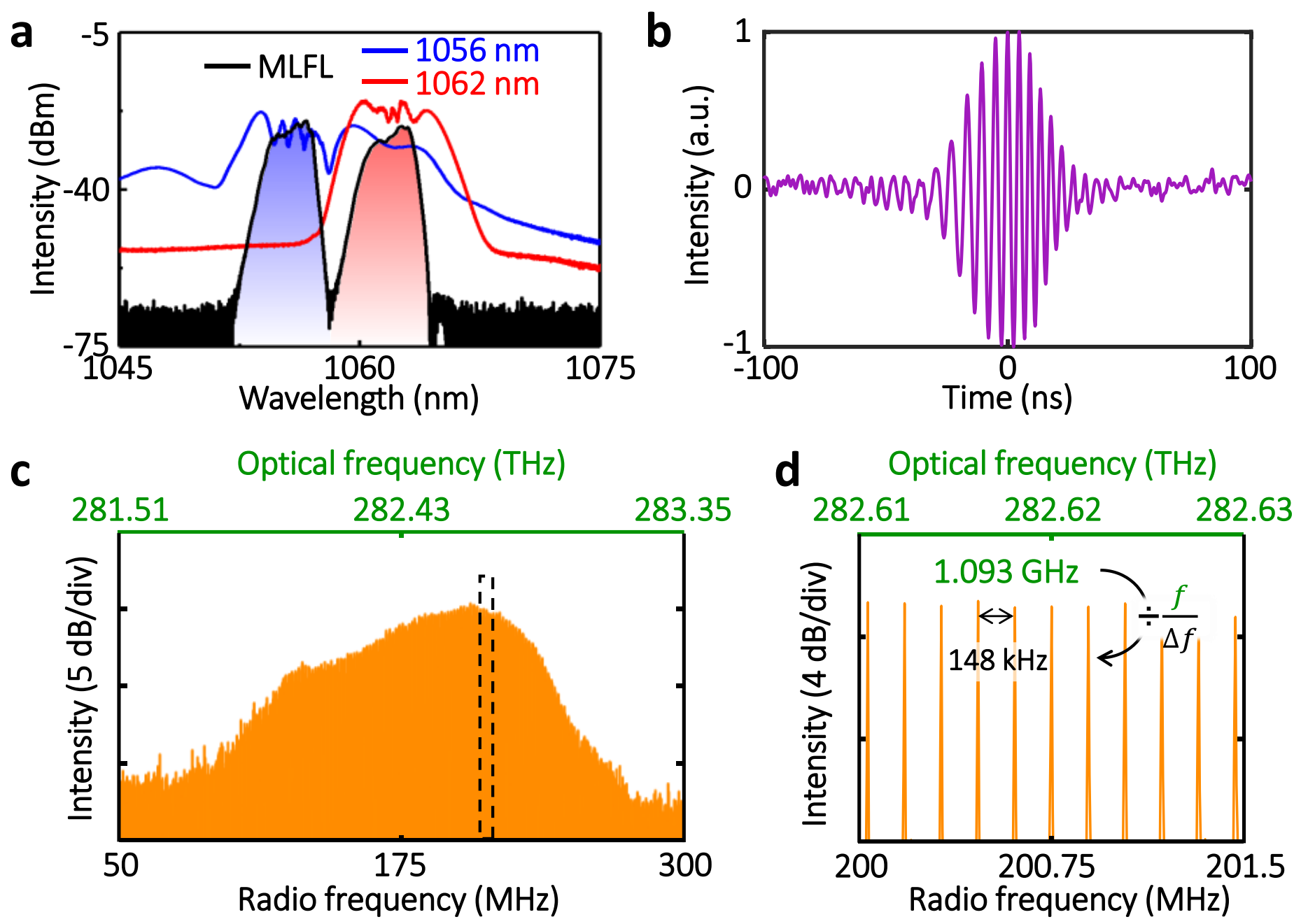

**Fig. 4 | Dual-comb generation through coherently spectral broadening. a.** Optical spectra of the dual-wavelength MLFL before and after coherently spectral broadening. **b.** Temporal interferogram of the DCS. **c.** Fourier-transformed RF spectrum. Here, the result is averaged over a time interval of 200 μs. **d.** Closeup of the comb tooth in **c**.

Coherently spectral broadening is then applied to each wavelength component for implementing the DCS (**Methods** and **Supplementary Note 7.1**). As shown in **Fig. 4a**, the optical spectra after broadening clearly show spectral overlap. The temporal interferograms of the dual-comb heterodyne signal are detected by the photodiode (PD) and recorded by a high-speed real-time oscilloscope, as shown in **Fig. 4b**, wherein the recorded signal is filtered by passing through a low-pass filter (**Supplementary Note 7.2**). The corresponding Fourier-transformed RF spectrum is displayed in **Fig. 4c,** and the details of the comb tooth can be observed from the closeup shown in **Fig. 4d**. Please note that, the nonlinear spectral broadening must be carefully manipulated, as inadequate spectral overlap will limit practical applications of the DCS, while excessive nonlinear accumulation will deteriorate the coherence property (**Supplementary Note 7.3**).

**High-speed measurements using GHz single-cavity all-fiber dual-comb laser**

To demonstrate its practical applications, we first evaluate the spectral accuracy and acquisition speed of the DCS. Fiber Bragg grating (FBG) is employed as the probe to sense the deformation that results in Bragg wavelength shifting. The DCS is performed to measure the change of Bragg wavelength and thus diagnose the strain applied to the FBG. **Figure 5a** presents the experimental setup, wherein the Fourier-transformed RF spectrum is reflected by two FBGs. FBG1 serves as the reference, and FBG2 is subjected to strain tuning by a piezoelectric (PZT, bottom right inset of **Fig. 5a**). The reflected signals from both FBGs are extracted by a fiber circulator (CIR). The original Bragg wavelengths of FBG1 and FBG2 are 1060.40 nm and 1061.01 nm, respectively. The temporal interferograms carrying Bragg wavelength information and the reflected spectra of FBGs are reconstructed through the DCS.

The spectral accuracy of the DCS is quantified by intuitively comparing the DCS measurement with a standard optical spectrum analyzer (OSA), as shown in **Fig. 5b** (also see **Supplementary Note 8.1**). Both the DCS and OSA measurements show that the frequency and wavelength changes exhibit good linearity as a function of PZT driving voltage. The ratio of frequency shift to voltage in the DCS measurement is estimated to be 112.13 kHz/V in the RF domain, corresponding to 828.08 MHz/V in the optical frequency domain. It is 3.11 pm/V (RF) and 829.79 MHz/V (optical frequency) in the OSA measurement, which closely matches that of the DCS measurement — demonstrating the capability of precise spectral measurement by using our GHz single-cavity all-fiber dual-comb laser.

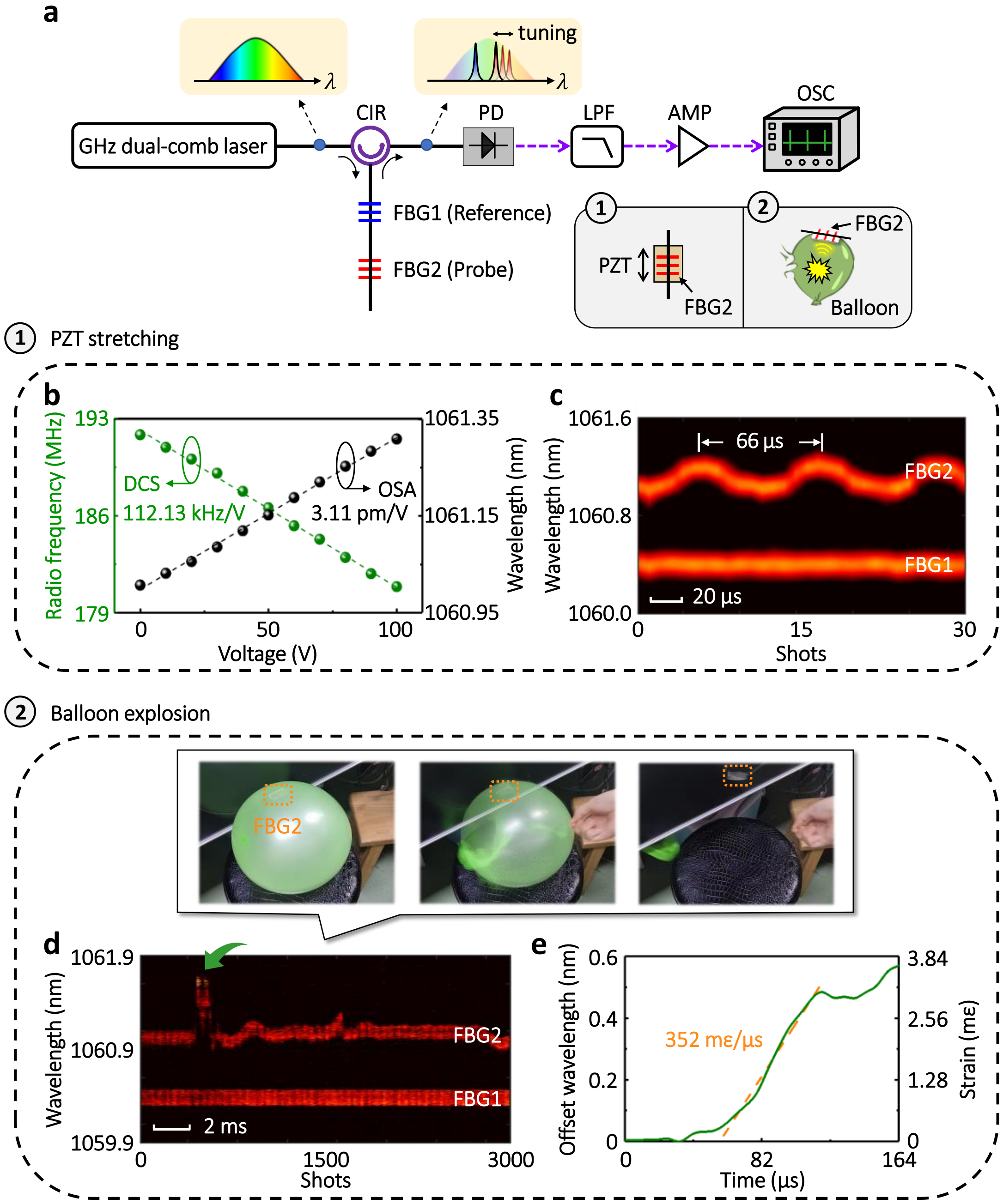

**Fig. 5 | Real-time DCS measurements of fast wavelength tuning and balloon explosion. a.** Experimental setup of high-speed DCS measurements. CIR, circulator; FBG, fiber Bragg grating; PZT, piezoelectric; PD, photodiode; LPF, low-pass filter; AMP, amplifier; OSC, oscilloscope. Two events are measured, i.e., PZT stretching (①) and balloon explosion (②). FBG2 is placed onto the PZT or balloon as the probe, while FBG1 serves as the reference. **b.** Frequency and wavelength changes as a function of PZT driving voltage, respectively measured by the DCS (green spots) and standard optical spectrum analyzer (OSA, black spots). **c.** Spectral evolution recorded by the DCS when the PZT stretches the FBG2 at a modulation frequency of 15 kHz. **d.** Spectral evolution recorded by the DCS during the balloon explosion. Insets show the snapshots of the balloon explosion experiment. FBG2 is placed on the air-filled balloon and covered by polydimethylsiloxane (PDMS). **e.** Wavelength and corresponding strain variation of FBG2 during the balloon explosion as indicated by the green arrow in **d**.

To demonstrate the high acquisition speed of the DCS, we capture the fast Bragg wavelength tuning of the FBG2 glued onto the PZT. A triangular signal with a modulation frequency of 15 kHz is applied to the PZT. The PZT imparts a dynamic strain to FBG2, thereby causing a rapid variation in its reflection wavelength. The corresponding spectral evolution of FBGs is presented in **Fig. 5c**, which involves 30 shots of DCS measurement over a time period of 200 μs. It is also noticed that the reflection wavelength of FBG1 remains unchanged, while FBG2 undergoes periodic changes consistent with the modulation frequency of PZT.

As a proof-of-concept DCS measurement of practical transient events, we capture the balloon explosion that can simulate the generation of 'shock wave'. In the experiment, the FBG2 is encapsulated with polydimethylsiloxane (PDMS, **Supplementary Note 9**), and then placed on the surface of an air-filled balloon. The balloon explosion is initiated by puncturing with a needle, causing a 'shock wave' (inset of **Fig. 5d**, also see **Supplementary Video 1**). The temporal interferograms carrying information about the variation of the reflection wavelength during the balloon explosion are demodulated by the Fourier transform, and the corresponding spectral evolution is shown in **Fig. 5d**. It can be observed that the intense 'shock wave' generated by the balloon explosion induces strain on the FBG2, resulting in a significant and rapid change to the reflection wavelength. **Figure 5e** depicts the relative wavelength change of FBG2 when the explosion occurs (indicated by the green arrow in **Fig. 5d**). The strain imparted on the FBG2 can be calculated from the transformed relationship (i.e., 1 nm = 6.41 mε, **Supplementary Note 8.2**), and the slope of linear fitting is about 352 mε/μs.

## Real-time DCS measurements outside the laboratory

To further showcase the practicality and reliability of the GHz single-cavity all-fiber dual-comb laser, we conduct real-time DCS measurements of firecracker explosions outside the laboratory. **Figure 6a** shows a photograph of the experimental implementation in the outdoor environment, wherein the dual-wavelength MLFL is engineeringly packaged (top left inset of **Fig. 6a**). As shown in the bottom left inset of **Fig. 6a**, FBG2 is adhered to the outer surface of a stainless steel container. When the firecracker explodes inside, the container's surface experiences a powerful shock, leading to a change in the reflection wavelength of FBG2 (**Supplementary Video 2**). In the experiment, containers with different thicknesses are employed, i.e., 1, 2, and 3 mm, and the corresponding spectral evolutions are recorded and presented in **Fig. 6b**. It is observed that the reflection wavelength of FBG2 undergoes a more significant and rapid change as the thickness of the container decreases, as illustrated in **Fig. 6c**. It implies that the FBG2 is subjected to a stronger strain for a thinner container, which is consistent with the intuitive understanding.

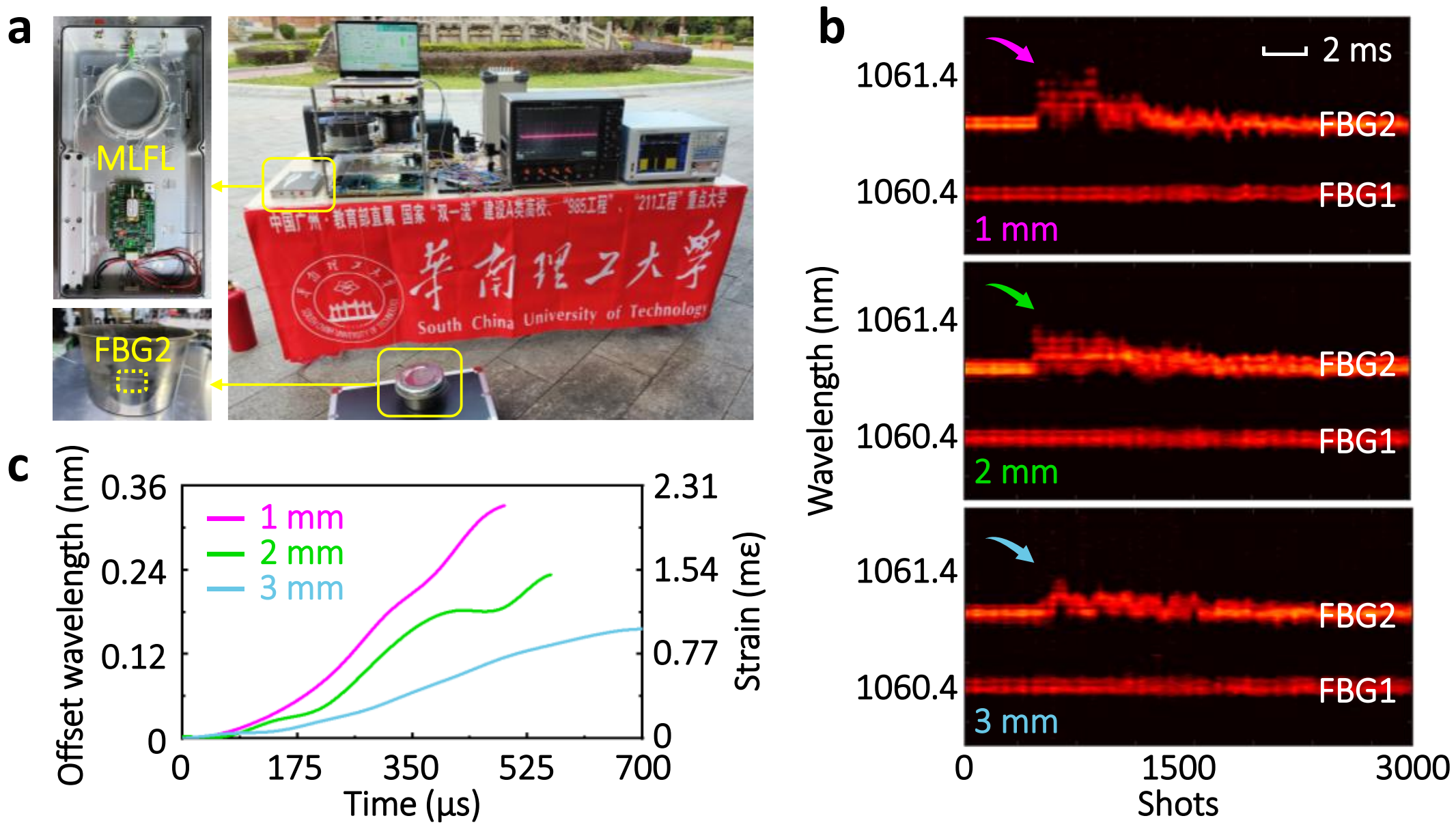

**Fig. 6 | Real-time DCS measurements of the firecracker explosion. a.** Photo of the measurement system outside the laboratory. Top left inset shows the engineering integration of dual-wavelength MLFL. Bottom left inset depicts the FBG2 adhered to the outer surface of the stainless steel container. **b.** Spectral evolution recorded by the DCS during the firecracker explosion inside the containers with different thicknesses, i.e., 1 mm (pink), 2 mm (green), and 3 mm (blue). **c.** Wavelength and corresponding strain variations during the firecracker explosions as indicated by the arrows in **b**.

## Conclusion

In summary, we have demonstrated a single-cavity dual-comb source generated from a dual-wavelength MLFL with GHz-level fundamental repetition rates. The asynchronous dichromatic pulse trains delivered from the ultrashort fiber cavity leveraging the MMI-mediated spectral filtering effect have a repetition rate difference of 148 kHz, which exhibits a low fluctuation with an Allan deviation of only 101.7 mHz for an average time of 1 second in the free-running state. The optical spectra of the asynchronous dichromatic pulses are then coherently broadened for the DCS generation enabling a short refresh time of 6.75 μs. Notably, this GHz DCS presents an appealing balance between spectral accuracy and acquisition speed. As proof-of-concept experiments, the GHz single-cavity all-fiber dual-comb system was applied to successfully record the nonrepeatable explosion events outside the laboratory, demonstrating its satisfactory practicality. It is anticipated that the present robust and compact dual-comb system will open new potential in various applications, especially for studying dynamic events, like shock waves, turbulences, and chemical reactions.

## Methods

**Experimental setup of the GHz single-cavity all-fiber dual-comb system**: The system mainly consists of a dual-wavelength MLFL with GHz-level fundamental repetition rates (**Supplementary Note 1**), optical amplifier and spectral-broadening unit (**Supplementary Note 7.1**). In the dual-wavelength MLFL, there involves an 8.5-cm heavily Yb-doped FMGF, a 0.8-cm SMF, an SESAM, and a fiber-based DF. The fiber-based DF is coated onto the fiber end facet in a ceramic ferrule with a signal reflectivity of 85%, and its pigtail is spliced to the common port of a 976/1064 nm wavelength-division multiplexer (WDM). A single-mode laser diode (SM-LD) serves as the pump source (centered at 976 nm with a maximum power of 460 mW), which is coupled into the cavity through the WDM. A 90:10 optical coupler (OC) splits the MLFL signal into two parts, with the 10% part used for monitoring. A fiber isolator (ISO) is adopted to protect the cavity from backward reflection. The asynchronous dichromatic pulse trains are separated by a filter wavelength-division multiplexer (FWDM). Each of the separated wavelength components is individually sent into a Yb-doped fiber amplifier (YDFA) to boost its average power to ~40 mW. Then, the amplified pulses are launched into highly nonlinear photonic crystal fiber (PCF, ~110 m in length) that has a nonlinear coefficient of 34 $W^{-1}km^{-1}$ for spectral broadening. Finally, the spectrally broadened pulse trains are combined through another OC.

**Numerical simulation**: In the simulation of generating asynchronous dichromatic pulse trains with GHz-level fundamental repetition rates, two nonlinear Schrödinger equations without the terms contributing to the cross-phase modulation and four-wave mixing are used, i.e.,

$$\frac{\partial u(z,t)}{\partial z} = -i\frac{\beta_2}{2}\frac{\partial^2 u(z,t)}{\partial t^2} + i\gamma|u(z,t)|^2 u(z,t) + \frac{g_u(z,\omega)}{2}u(z,t), \tag{2a}$$

$$\frac{\partial v(z,t)}{\partial z} = -i\frac{\beta_2}{2}\frac{\partial^2 v(z,t)}{\partial t^2} + i\gamma|v(z,t)|^2 v(z,t) + \frac{g_v(z,\omega)}{2}v(z,t), \tag{2b}$$

where $u$ and $v$ are the field envelopes of the dichromatic pulses. $\beta_2$ and $\gamma$ represent the second-order dispersion and nonlinearity of the fiber, respectively. The gain coefficients $g_u$ and $g_v$ are calculated by the propagation-rate equations that are respectively associated with field $u$ and $v$, while the gain saturation resulted from the superposition of dichromatic pulses is not involved. To consider the interaction between the two pulses, the saturable absorption $q(t)$ of the SESAM is depicted by

$$\frac{dq}{dt} = -\frac{q(t)-q_0}{T_a} - \frac{q(t)}{E_a}\left(\left|u\left(L_f^-,t\right)\right|^2 + \left|v\left(L_f^-,t\right)\right|^2\right), \tag{3}$$

where $q_0$, $T_a$, $E_a$ are the modulation depth, relaxation time, and saturation energy of the SESAM, respectively. $L_f$ is the fiber length. The propagation of the high-order transverse mode in the FMGF is equivalently considered by the employment of a pairwise artificial filters, i.e.,

$$for\ u, \qquad T_u = exp\left(-\left(\frac{\omega - \omega_{offset}}{\Delta\omega}\right)^8\right), \tag{4a}$$

$$for\ v, \qquad T_v = exp\left(-\left(\frac{\omega + \omega_{offset}}{\Delta\omega}\right)^8\right), \tag{4b}$$

where $2\omega_{offset}$ designates the wavelength spacing of the MMI-mediated spectral filter, and $\Delta\omega$ indicates the bandwidth of the filter. Key parameters used in the simulation include $\beta_2$ = 30 $fs^2$/mm, $\gamma$ = 4.5 $W^{-1}km^{-1}$, $q_0$ = 0.05, $T_a$ = 1 ps, $E_a$ = 5.7 pJ, $\omega_{offset}$ = 5.3 THz, and $\Delta\omega$ = 3 THz. More details about the numerical simulation are provided in **Supplementary Note 4**.

**Data acquisition**: The output power of the dual-wavelength MLFL is monitored by a photodiode-based power meter (Thorlabs PM100D & S122C), while the optical spectrum is analyzed by an OSA (YOKOGAWA AQ6370D). The optical pulses are converted to electric signals via a high-speed PD (Newport 818-BB-51F, 12.5 GHz bandwidth), and recorded by a standard real-time oscilloscope (Teledyne LeCroy WaveMaster 820Zi-B, 20 GHz bandwidth). The RF performance, including phase noise and intensity noise, is analyzed by a frequency signal analyzer (Rohde & Schwarz FSWP26, 26.5 GHz bandwidth). A frequency counter (Keysight, 53220A) is used to evaluate the long-term frequency stability. The autocorrelation trace of the mode-locked pulse is measured by an autocorrelator (Femtochrome, FR-103XL).

**Data availability**

All data used in this study are available from the corresponding authors upon reasonable request.

**Code availability**

All custom codes used in this study are available from the corresponding authors upon reasonable request.

**Acknowledgments**

This work was partially supported by National Natural Science Foundation of China (62375087 and 12374304), Key-Area Research and Development Program of Guangdong Province (2023B0909010002), NSFC Development of National Major Scientific Research Instrument (61927816), Mobility Programme of the Sino-German (M-0296), Introduced Innovative Team Project of Guangdong Pearl River Talents Program (2021ZT09Z109), and Natural Science Foundation of Guangdong Province (2021B1515020074).

**Author contributions**

L.L., W.L. and Z.H.L. performed the experiments. L.L., W.L., Z.H.L, M.J.P, C.Y.W, Y.Y., Y.K.G, and Z.J.X. conducted the theoretical analysis and numerical simulations. L.L., W.L., and X.W.C. processed and analyzed the data. L.L. and X.M.W. wrote the manuscript. All authors commented on the manuscript. X.M.W. and Z.M.Y. supervised the project.

**Competing financial interests**

The authors declare no competing interests.